\documentclass[apj]{emulateapj}

\usepackage{graphicx,times}
\usepackage{subfigure}
\newcommand{\be}{\begin{equation}}
\usepackage{threeparttable}
\usepackage{booktabs}
\newcommand{\ee}{\end{equation}}
\newcommand{\bea}{\begin{eqnarray}}
\newcommand{\eea}{\end{eqnarray}}

\usepackage{enumerate}
\usepackage{amsmath}
\usepackage{cases}
\usepackage{longtable}
\usepackage{hyperref}
\usepackage{epstopdf}
\usepackage{amsmath,bm}
\usepackage{amssymb}
\usepackage{natbib}
\usepackage{morefloats}
\usepackage{multirow}
\usepackage{array}
\usepackage{verbatim}

\begin{document}

\title{On the formation of density filaments in the turbulent interstellar medium}

\author{Siyao Xu\altaffilmark{1,2} and Alex Lazarian\altaffilmark{1} }

\altaffiltext{1}{Department of Astronomy, University of Wisconsin, 475 North Charter Street, Madison, WI 53706, USA; 
sxu93@wisc.edu,
lazarian@astro.wisc.edu}
\altaffiltext{2}{Hubble Fellow}

\begin{abstract}

This study is motivated by recent observations on the ubiquitous interstellar density filaments and 
guided by the modern theory of magnetohydrodynamic (MHD) turbulence. 
The interstellar turbulence shapes the observed density structure. 
The perpendicular turbulent mixing, as the fundamental dynamics of MHD turbulence, naturally entails
parallel filaments presented in both the diffuse medium and molecular clouds (MCs).
The minimum width is determined by the perpendicular neutral-ion decoupling scale in partially ionized media. 
Differently, dense perpendicular filaments arise in highly supersonic turbulence in MCs as a result of 
shock compression. Their width specifically depends on the turbulence properties. 
We demonstrate that different alignments of filaments with respect to the magnetic field originate from the varying 
turbulence properties in the multi-phase interstellar medium.

\end{abstract}


\section{Introduction}

Observations reveal that filamentary density structures are widely spread in the interstellar medium (ISM), 
including both the diffuse neutral medium 
(e.g., \citealt{McC06,Cl14,Pla16})
and highly fragmented molecular clouds (MCs)
(e.g., \citealt{SE79,Will00,And10}).
Moreover, the filaments in the diffuse medium preferentially align with the local magnetic field, 
whereas the dense filaments in MCs tend to be perpendicular
\citep{Pla16,Pln16}.
The former provides the information on the Galactic interstellar magnetic field. The latter is important for understanding the star formation process.
as dense filaments in MCs coincide with the birthplaces
of protostellar cores
\citep{And14,Mar16}.
Besides its significance in interstellar processes, aligned with magnetic field filaments in HI were shown to be useful for  
studies of cosmic microwave background polarization
\citep{Cl15}.

As the ISM is turbulent
\citep{Armstrong95,CheL10}, 
understanding the turbulence properties 
is essential for explaining the magnetic field and density structure in the multi-phase ISM. 
In the high-latitude diffuse neutral medium, the turbulence is injected at $\sim 140$ pc and mildly supersonic 
\citep{Chep10},
while the turbulence in MCs in the Galactic plane is highly supersonic and shock-dominated, with sonic Mach numbers $M_s \sim 5-20$
\citep{Zu74,Zuck74,Lars81}.
In accordance with the distinctive turbulence properties,
we naturally expect the resulting density structure differs in different media.

Important strides have been made towards advancing the theories of magnetohydrodynamic (MHD) turbulence 
concerning e.g. the dynamics and statistics of MHD turbulence
(\citealt{GS95}, hereafter GS95; \citealt{LV99}, hereafter LV99),
MHD turbulence in compressible medium
\citep{LG01,CL02_PRL,CL03}
and in partially ionized medium 
\citep{LG01,LVC04,XuL15,Xuc16},
and their numerical testing 
\citep{MG01,CLV_incomp,KL09,KL12}.
With a general applicability in essentially all branches of astrophysics involving fluid dynamics 
\citep{Mckee_Ostriker2007},
the theoretical developments of MHD turbulence provide new insights in many long-standing problems 
\citep{YL02,XuZ17,XZg17}.
In particular, the turbulent reconnection of magnetic fields inherent in MHD turbulence 
(LV99)
was introduced to solve the magnetic flux problem in star formation
\citep{Laz05,Sant10, LEC12}, 
and the turbulence anisotropy has been employed to develop a
new velocity gradient technique for measuring interstellar magnetic fields 
\citep{Gon17,Yu17, LY18}.

The interstellar turbulence induces density fluctuations and influences the density structure. 
Statistical studies 
\citep{BLC05,KL07,Lad08,Burk09,Burk10, Col12,Fed12, Bur15}
uncover different density structures developed in various turbulence regimes. 
Density fluctuations in compressible MHD turbulence are passively regulated by the Alfv\'{e}nic turbulence, 
which is responsible for the dynamics of MHD turbulence, 
and thus present the same statistical feature as turbulent velocities. 
When the compressible turbulence becomes highly supersonic, 
besides the density fluctuations associated with the Alfv\'{e}n component, 
the shocks driven by supersonic turbulent flows can also produce additional density structure, 
which is characterized by large density contrasts at small scales due to the shock compression. 
The fact of different density structures arising in distinctive turbulence regimes in diverse ISM phases 
has been shown by overwhelming observations
(see e.g., \citealt{Armstrong95,CheL10}; reviews by \citealt{Laz09rev,HF12} and references therein),
which has been applied by
\citet{XuZ17}
to modeling the Galactic electron density fluctuations and explaining the scattering measurements of pulsars.

In this paper, based on the established and tested theory of MHD turbulence, 
we will elucidate the origin of density filaments in the turbulent and magnetized ISM, 
including both parallel filaments that generally arise not only in the diffuse neutral medium but also in MCs (Section 2),
and dense perpendicular filaments in highly supersonic MCs
(Section 3). 
The discussion and summary are in Sections 4 and 5.

\section{Filaments parallel to the magnetic field}

In the compressible ISM, 
the dynamics of MHD turbulence is dominated by its Alfv\'{e}nic component
\citep{CL02_PRL,CL03}.
The density structure shaped by the mixing motions of Alfv\'{e}nic turbulence generally 
exists in both the diffuse ISM and MCs.

\subsection{Perpendicular turbulent mixing and parallel filaments}

The turbulent energy injected at a large scale, i.e., the injection scale $L$, 
progressively cascades down to small scales. 
In MHD turbulence, the turbulent mixing 
of magnetic fields takes place in the direction perpendicular to the {\it local} magnetic field,
which is enabled by the fast turbulent reconnection (LV99).
This feature renders the turbulence anisotropy and 
makes the essential distinction between the MHD and hydrodynamic turbulence.
The perpendicular cascade of energy is more rapid than the parallel one. 
In consequence, the turbulent energy is distributed anisotropically, with most turbulent energy residing 
in the perpendicular direction.

The nonlinear cascade rate of Alfv\'{e}nic turbulence, i.e., the eddy-turnover rate, 
determines the turbulent mixing rate, 
\begin{equation}\label{eq: velgs}
  \tau_\text{cas}^{-1} = v_l l_\perp^{-1} = V_\text{st} L_\text{st}^{-\frac{1}{3}}  l_\perp^{-\frac{2}{3}}  ,
\end{equation}
with the local turbulent velocity $v_l$ at the length scale measured in the direction 
perpendicular to the local magnetic field $l_\perp$. 
The critical balance between the rate of this perpendicular mixing and 
the frequency of Alfv\'{e}n waves propagating along the magnetic field is satisfied in the strong MHD turbulence regime 
(GS95).
In the above expression, the GS95 scaling for the strong MHD turbulence is used: 
\begin{equation}\label{eq: gsloc}
   v_l = V_\text{st} (l_\perp / L_\text{st})^{\frac{1}{3}} , 
\end{equation}
where $V_\text{st}$ is the turbulent velocity at the injection scale of the strong MHD turbulence $L_\text{st}$. 
More specifically, in super-Alfv\'{e}nic turbulence with dominant turbulent kinetic energy at the driving scale, there is 
\begin{equation}\label{eq: supalf}
   V_\text{st} = V_A,  L_\text{st} = L M_A^{-3} , 
\end{equation}
where $V_A$ is the Alfv\'{e}n speed, 
$M_A = V_L / V_A >1$ is the Alfv\'{e}n Mach number, and $V_L$ is the turbulent velocity at $L$. 
Within the range of length scales $[L, L_\text{st}]$, the turbulence is in the hydrodynamic regime with isotropic turbulent mixing and 
turbulent energy distribution. 
Differently, in the case of sub-Alfv\'{e}nic turbulence with dominant magnetic energy at the driving scale, i.e., $M_A <1$, there is
\citep{Lazarian06}, 
\begin{equation}\label{eq: subalf}
   V_\text{st} = V_L M_A, L_\text{st} = L M_A^2   .
\end{equation}
Within $[L, L_\text{st}]$, it is the weak turbulence, i.e., weakly interacting Alfv\'{e}n waves
(LV99).

Given the expression of the perpendicular mixing rate (Eq. \eqref{eq: velgs}), the critical balance condition
\begin{equation}\label{eq: criba}
    \tau_\text{cas}^{-1} = V_A / l_\|
\end{equation}
leads to the anisotropic scaling relation of the strong MHD turbulence, 
\begin{equation}\label{eq: turani}
   l_\| = \frac{V_A}{V_\text{st}} L_\text{st}^{\frac{1}{3}} l_\perp^{\frac{2}{3}},
\end{equation}
where $l_\|$ is the parallel length scale of a turbulent eddy.  
It shows that the anisotropy of the GS95 turbulence is scale-dependent, with 
smaller-scale turbulent eddies more elongated along the local magnetic field.

The turbulent mixing governs the mixing of density fluctuations. 
Density fluctuations in compressible MHD turbulence come from the 
slow mode and the entropy mode
\citep{LG01}.
The resulting density inhomogeneities behave as a passive scalar and 
are passively mixed by Alfv\'{e}nic turbulence, following the same perpendicular cascade as turbulent velocities. 
As a result, similar to the anisotropic turbulent velocity eddies,
they tend to exhibit an elongated structure along the local magnetic field, i.e., a parallel filament. 
The lifetime of the parallel filament is determined by the turnover time of the turbulent eddy at the perpendicular length scale equal 
to the filament width. 
In the compressible MHD turbulence and even at a high $M_s = V_L/c_s$, where $c_s$ is the sound speed,
the Alfv\'{e}n modes carry most of the turbulent kinetic energy, and the mixing motions induced 
by the Alfv\'{e}nic turbulence are marginally affected by the compressible modes
(see e.g. \citealt{CL02_PRL,CL03,Pan10}).
This explains the existence of parallel filaments in supersonic MHD turbulence.

\subsection{Neutral-ion decoupling and the minimum width of parallel filaments}

The ISM is commonly partially ionized 
\citep{Draine:2011aa}.
The coupling between neutrals and ions should be taken into account when we study the 
density structure in neutral gas. 
The coupling state between neutrals and ions depends on the length scale of interest. 
The parallel neutral-ion decoupling scale is defined as 
\begin{equation}\label{eq: panid}
     l_{\text{ni, dec}, \|} = \frac{V_A}{\nu_{ni}}, 
\end{equation}
corresponding to the scale at which the Alfv\'{e}n wave frequency and 
the neutral-ion collision frequency $\nu_{ni}= \gamma_d \rho_i$ get equal, 
where $\gamma_d$ is the drag coefficient 
(see e.g. \citealt{Shu92}), 
and $\rho_i$ is the ion mass density. 
According to the anisotropic scaling of Alfv\'{e}nic turbulence (Eq. \eqref{eq: turani}), 
the perpendicular neutral-ion decoupling scale is, 
\begin{equation}\label{eq: genkdec}
  l_{\text{ni, dec}, \perp} = \nu_{ni}^{-\frac{3}{2}} L_\text{st}^{-\frac{1}{2}} V_\text{st}^{\frac{3}{2}}. 
\end{equation}
In the strong MHD turbulence regime, 
over large scales above the perpendicular decoupling scale, the strongly coupled neutrals and ions together carry the 
anisotropic MHD turbulence, where density fluctuations present a filamentary structure along the local magnetic field 
as a result of the perpendicular turbulent mixing. 
Over smaller scales, 
neutrals decouple from ions, as well as the magnetic field, and they independently carry the isotropic hydrodynamic turbulence
\citep{XuL15, Burk15}. 
The MHD cascade in neutrals terminates at $l_{\text{ni, dec}, \perp}$, which is thus the minimum width of 
parallel filaments in density fluctuations. 

To numerically illustrate the width range of parallel filaments in the partially ionized ISM, we adopt the typical driving conditions of the interstellar 
turbulence, $L = 30$ pc, $V_L = 10$ km s$^{-1}$, 
and typical conditions of the warm neutral medium (WNM), 
the cold neutral medium (CNM), and molecular clouds (MCs)
\citep{Dr98} (as listed in Table \ref{tab: cnm}). 
Besides, we use $\gamma_d=3.5\times10^{13}$cm$^3$g$^{-1}$s$^{-1}$ 
\citep{Drai83},
the ion and neutral masses $m_i = m_n = m_H$ for the WNM and CNM, 
and $m_i = 29 m_H$, $m_n = 2.3 m_H$ for MCs 
\citep{Shu92}, 
where $m_H$ is the mass of hydrogen atom.

In the case of the WNM, given the above parameters, the turbulence is sub-Alfv\'{e}nic. 
Inserting Eq. \eqref{eq: subalf} in Eq. \eqref{eq: genkdec} yields 
\begin{equation}\label{eq: subkde}
    l_{\text{ni, dec}, \perp, \text{sub}} = \nu_{ni}^{-\frac{3}{2}} L^{-\frac{1}{2}} V_L^{\frac{3}{2}} M_A^{\frac{1}{2}}. 
\end{equation}
Its value is presented in Table \ref{tab: cnm}. 
To quantify the elongation of the decoupling-scale filament, we also calculate the axial ratio as (Eqs. \eqref{eq: panid} and \eqref{eq: subkde})
\begin{equation}
      A_\text{dec, sub} = \frac{l_{\text{ni, dec}, \|}}{l_{\text{ni, dec}, \perp, \text{sub}}} = \nu_{ni}^\frac{1}{2} L^\frac{1}{2} V_L^{-\frac{1}{2}} M_A^{-\frac{3}{2}},
\end{equation}
which has a large value in the sub-Alfv\'{e}nic WNM. 
In cases of the CNM and MCs, with the above parameters adopted, the turbulence turns out to be super-Alfv\'{e}nic. 
We can rewrite Eq. \eqref{eq: genkdec} by using Eq. \eqref{eq: supalf},
\begin{equation}\label{eq: desups}
  l_{\text{ni, dec}, \perp, \text{sup}} = \nu_{ni}^{-\frac{3}{2}} L^{-\frac{1}{2}} V_L^{\frac{3}{2}}  ,
\end{equation}
which is independent of the magnetic field strength, unlike the case in sub-Alfv\'{e}nic turbulence.
The axial ratio at the decoupling scale is 
(Eqs. \eqref{eq: panid} and \eqref{eq: desups})
\begin{equation}
     A_\text{dec, sup} = \frac{l_{\text{ni, dec}, \|}}{l_{\text{ni, dec}, \perp, \text{sup}}} = \nu_{ni}^\frac{1}{2} L^\frac{1}{2} V_L^{-\frac{1}{2}} M_A^{-1}.
\end{equation}
The reason for the small $A_\text{dec, sup}$ in the CNM and MCs is that 
with the decoupling scale close to $L_\text{st}$, the turbulence anisotropy at the decoupling scale is insignificant. 
Moreover, in super-Alfv\'{e}nic turbulence, as the anisotropic turbulent mixing only operates below $L_\text{st}$, 
the upper limit to the width of a parallel filament is given by $L_\text{st}$.

From the above analysis, we find that 
in both the transonic WNM and supersonic CNM and MCs
(see $M_s$ values in Table \ref{tab: cnm}), the perpendicular turbulent mixing caused by the 
Alfv\'{e}nic turbulence gives rise to density filaments parallel to the local magnetic field. 
Furthermore, 
comparing the sub-Alfv\'{e}nic WNM and the super-Alfv\'{e}nic CNM and MCs, we see that the former is more favorable for the formation of 
profoundly elongated parallel filaments.

\begin{table*}[t]
\centering
\begin{threeparttable}
\caption[]{Parameters in different ISM phases}\label{tab: cnm} 
  \begin{tabular}{cccccccccc}
     \toprule
              &  $n_H [\text{cm}^{-3}]$ &  $n_e/n_H$         & $B_0$ [$\mu$ G] &  $T$ [K]  & $M_s$    & $M_A$    &  $L_\text{st}$ [pc]    &  $l_{\text{ni, dec}, \perp}$ [pc]  & $A_\text{dec}$  \\
      \hline
   WNM  & $0.4$                          &  $0.1$                & $5$           &    $6000$  &  $1.1$         &  $0.6$      &   $11.1$                         &    $7.3\times10^{-3}$                   &  $31.1$     \\
   CNM  &  $30$                           &  $10^{-3}$         &  $5$           &    $100$    &  $8.5$         &     $5$       &   $0.2$                          &    $1.5 \times10^{-2}$                  &  $2.5$    \\
   MC     &  $300$                         &  $10^{-4}$         & $5$            &    $20$     &   $28.9$       &     $15.9$   &   $7.5\times10^{-3}$     &  $9.3\times10^{-5}$                     &  $4.3$   \\
    \bottomrule
    \end{tabular}
 \end{threeparttable}
\end{table*}

We note that the calculations here only serve as illustrative examples. To explain specific observations, more realistic parameters 
depending on the local environments should be used.

\section{Filaments perpendicular to the magnetic field}

In MCs, in addition to the parallel filaments formed 
in the Alfv\'{e}nic turbulence, 
the shock compression in highly supersonic turbulence generates 
dense perpendicular filaments.

\subsection{Shock compression in highly supersonic turbulence}

We consider a shock wave 
driven by the supersonic turbulent flow propagating in MCs. 
The initial shock velocity is given by the turbulent velocity $V_L$ at the driving length scale $L$ of the supersonic turbulent flow. 
The corresponding turbulent sonic Mach number is 
$M_{sL} = V_L / c_s$.

In the rest frame of the shock, the quantities across the shock are related by the jump conditions, which include the 
conservation of mass, 
\begin{equation}\label{eq: conm}
    \rho_1 u_1  = \rho_2 u_2,
\end{equation}
the conservation of momentum, 
\begin{equation}\label{eq: csmo}
    \rho_1 c_1^2 + \rho_1 u_1^2 + \frac{B_1^2}{8\pi} =  \rho_2 c_2^2 + \rho_2 u_2^2 + \frac{B_2^2}{8\pi},
\end{equation}
and the conservation of magnetic flux 
\begin{equation}\label{eq: csfl}
       B_1 u_1 =  B_2 u_2, 
\end{equation}
where $\rho_1$, $u_1$, $c_1$, $B_1$ are the mass density, fluid velocity (in the shock propagation direction),
the sound speed, the strength of the transverse magnetic field in the upstream region, 
and $\rho_2$, $u_2$, $c_2$, $B_2$ are those in the downstream region.

Under the consideration of highly supersonic turbulence, Eq. \eqref{eq: csmo} can be approximately written as 
\begin{equation}
   \rho_1 u_1^2 + \frac{B_1^2}{8\pi}  \approx  \rho_2 c_s^2 + \rho_2 u_2^2 + \frac{B_2^2}{8\pi}.
\end{equation}
Here we also assume that the shock is isothermal with $c_1 = c_2 = c_s$ due to the efficient cooling in MCs 
(see e.g. \citealt{Wh97}).
In combination with Eqs. \eqref{eq: conm} and \eqref{eq: csfl}, the above equation becomes
\begin{equation}
      \Bigg(1-\frac{V_{A2}^2}{2u_1^2} \Bigg) u_2^2  - u_1 u_2 +  c_s^2 + \frac{V_{A2}^2}{2}  \approx 0,
\end{equation}
as a quadratic equation of $u_2$, 
where $V_{A2} = B_2 /\sqrt{4\pi\rho_2}$. 
It has the solutions: 
\begin{equation}\label{eq:sol}
     u_2 \approx \frac{u_1 \pm u_1\sqrt{1 - 4 \Big(1-\frac{V_{A2}^2}{2 u_1^2}\Big) \Big(\frac{c_s^2}{u_1^2} + \frac{V_{A2}^2}{2 u_1^2} \Big)}}{2 - \frac{V_{A2}^2}{u_1^2}}  .
\end{equation}

In the case of dominant magnetic pressure in the downstream medium, i.e., $ c_s^2 \ll  V_{A2}^2 $, 
Eq. \eqref{eq:sol} can be simplified as 
\begin{equation}
    u_2 \approx \frac{u_1 \pm u_1 \Big(1-\frac{V_{A2}^2}{u_1^2}\Big)}{2},
\end{equation}
where $V_{A2}^2/u_1^2 \ll 1$ should be satisfied. 
We consider the solution corresponding to non-negligible compression of the shocked material and thus obtain, 
\begin{equation}\label{eq: uudrd}
    u_2 \approx \frac{V_{A2}^2}{2u_1}.
\end{equation}
From Eqs. \eqref{eq: conm} and \eqref{eq: csfl}, we find $V_{A2}^2 = (\rho_2/\rho_1) V_{A1}^2$, where $V_{A1} = B_1 /\sqrt{4\pi\rho_1}$,
and that the relation in Eq. \eqref{eq: uudrd} determines the density contrast across the shock as
\begin{equation}\label{eq:crsal}
    \frac{\rho_2}{\rho_1} \approx \sqrt{2} \frac{u_1}{V_{A1}} = \sqrt{2} M_{A1},
\end{equation}
where $M_{A1}$ is the shock Alfv\'{e}n Mach number. 
Evidently, to have a large density enhancement behind the shock, 
the transverse magnetic field in the upstream medium should be sufficiently weak.

In the case of dominant thermal pressure in the downstream medium, i.e., $c_s^2 \gg  V_{A2}^2$, the solutions in Eq. \eqref{eq:sol} can be reduced to 
\begin{equation}
      u_2 \approx \frac{u_1 \pm u_1 \sqrt{1- \frac{4 c_s^2}{u_1^2}}}{2}. 
\end{equation}
We again only consider the situation with a significant compression and find
\begin{equation}
     u_2 \approx \frac{c_s^2}{u_1}
\end{equation}
for a supersonic $u_1$. 
Combining the above expression with Eq. \eqref{eq: conm} yields
\begin{equation}\label{eq:crsm}
    \frac{\rho_2}{\rho_1} \approx \frac{u_1^2}{c_s^2}  = M_{s1}^2,
\end{equation}
where $M_{s1}$ is the shock sonic Mach number.

The above approximate compression ratios (Eqs. \eqref{eq:crsal} and \eqref{eq:crsm})
are consistent with earlier studies on isothermal shocks 
(see, e.g., \citealt{Draine:2011aa}). 
In both cases, a weak transverse magnetic field is required to facilitate a substantial density increase across the shock. 
It implies that in highly supersonic turbulence, the effective shock compression preferentially takes place in 
quasi-parallel shocks where the magnetic field is nearly aligned with the shock normal.
Moreover, based on the comparison between the two cases, 
with a quadratic dependence on $M_{s1}$, we expect a stronger shock compression when the downstream thermal pressure dominates 
over the magnetic pressure.

\subsection{Formation of perpendicular filaments}

To analyze the most prominent density structure 
resulting from the shock compression in highly supersonic turbulence,
we next focus on the case of a quasi-parallel shock with a negligible magnetic pressure in the downstream medium.

In the observer's frame, the downstream fluid velocity is (Eqs. \eqref{eq: conm} and \eqref{eq:crsm})
\begin{equation}
       v_2 = V_{so} - u_2 = V_{so} \Big(1 - \frac{1}{M_{so}^2}\Big),
\end{equation}
which is comparable to the shock velocity $V_{so} (= u_1)$ when the shock sonic Mach number
$M_{so} (= M_{s1})$ is large. 
Thus the swept-up material by the shock 
is compressed into a dense and thin sheet behind the shock front and co-moves with the shock. 
As a simple model, we suppose a planar shock wave with the swept-up mass concentrated in the post-shock sheet, 
\begin{equation}\label{eq:shem}
     M_\text{se} = R A \rho_1 =\delta A \rho_\text{se}.
\end{equation}
Here $R$ is the distance that the shock front has moved through.
$M_\text{se}$, $\delta$, $A$, and $\rho_\text{se}$ are the mass, thickness, area, and volume density of the sheet.

During the propagation of the shock wave, the momentum of the sheet $P_\text{se}$ is conserved, 
\begin{equation}
     P_\text{se} \approx M_\text{se} V_\text{so} = A \rho_1 R \frac{dR}{dt} = C, 
\end{equation}
where $C$ represents a constant. 
It yields
\begin{equation}
    R(t) = \sqrt{\frac{2C}{A\rho_1}} t^\frac{1}{2}
\end{equation}
and 
\begin{equation}
    V_\text{so} (t) =  \frac{1}{2}  \sqrt{\frac{2C}{A\rho_1}}  t^{-\frac{1}{2}}
\end{equation}
for $t>0$. The accumulation of the upstream ISM slows down the supersonic turbulent flow and weakens the shock. 
Accordingly, the compression ratio (Eq. \eqref{eq:crsm}) decreases with time, 
\begin{equation}
    M_\text{so}^2 (t) =  \frac{V_\text{so}^2 (t)}{c_s^2} = \frac{C}{2 A \rho_1 c_s^2} t^{-1}.
\end{equation}
Its time dependence shows that the large density contrast between the post-shock sheet and the ambient medium 
is mainly produced by the shock compression at a early time. 
Therefore, by assuming a constant compression ratio determined by the initial $M_{so} (= M_{sL})$ and $R$ 
given by the driving scale $L$ of the supersonic turbulent flow, we can approximately have the sheet thickness
(Eq. \eqref{eq:shem}):
\begin{equation}\label{eq:shtn}
   \delta \sim \frac{L}{M_{sL}^2},
\end{equation}
which only depends on the turbulence parameters. 
Provided $L \sim 10$ pc and $M_{sL} \sim 10$, 
we can estimate that $\delta$ is of order $0.1$ pc.

The dense sheet formed in a quasi-parallel shock is threaded by perpendicular magnetic fields. 
As multiple quasi-parallel shocks are generated in highly supersonic turbulence, 
their interactions naturally lead to the intersections of sheets
\citep{Ban06,Pud13},
where filaments form in the direction 
perpendicular to the surrounding magnetic fields. 
The filament width is expected to be comparable to the sheet thickness (Eq. \eqref{eq:shtn}). 
It varies at different driving conditions and compressibilities of turbulence.

\section{Discussion}

High-density filaments set up the necessary condition for the self-gravity to take over the 
gas dynamics and initiate the subsequent star formation. 
Such dense filaments can be easily generated by the shock compression in highly supersonic turbulence and tend to be 
perpendicular to the local magnetic field. 
As a different mechanism to produce filaments in turbulence, 
the turbulent mixing acts to dilute the density contrast and leads to low-density parallel filaments.
If the perpendicular filaments are not gravitationally bound 
and the filament width is within the inertial range of Alfv\'{e}nic turbulence, 
they are also subject to the turbulent mixing effect.

In the weak turbulence regime of sub-Alfv\'{e}nic turbulence, the turbulence is anisotropic with only perpendicular cascade of energy
(LV99). 
Despite the absence of turbulent mixing, density fluctuations can still present the parallel filamentary structure following the perpendicular cascade.

Theoretically, the width of parallel filaments can exist over the range $[L, l_{\text{ni, dec}, \perp, \text{sub}}]$ in sub-Alfv\'{e}nic turbulence and 
$[L_\text{st}, l_{\text{ni, dec}, \perp, \text{sup}}]$ in super-Alfv\'{e}nic turbulence.
Observationally, 
because smaller turbulent eddies are more numerous, we are more likely to 
identify the filaments at the smallest resolvable scale, which is usually larger than 
the minimum filament width. 
If the resolution is even larger than $L_\text{st}$ in super-Alfv\'{e}nic turbulence, 
it is unlikely to discern the mixing-induced parallel filaments. 
Besides, observations are also subject to the projection effect. 
The observed filament can actually be the superposition of distinct filaments along the line of sight.

For a filamentary density structure, the density gradient is perpendicular to its major axis.
Accordingly, the density gradient of a parallel filament is perpendicular to the magnetic field ${\bf B}$, 
while that of a perpendicular filament is parallel to ${\bf B}$. 
In a good agreement with our analysis, 
\citet{Yu17}
found that the density gradients measured in the diffuse HI region are perpendicular to ${\bf B}$, 
and 
\citet{YuL17}
showed density gradients parallel to ${\bf B}$ at shocks. 
Here we caution that the density gradient is not necessarily an indicator of a density filament. 
For instance, density gradients parallel to ${\bf B}$ can also arise in a gravitationally collapsing region. 
Moreover, under the effect of self-gravity, 
a perpendicular filament can experience a supercritical collapse, with the interior magnetic field dragged by the 
longitudinal infall and parallel to the major axis. 
\footnote{This change of magnetic field orientations can cause turbulent magnetic reconnection, which in turn removes 
the magnetic reversals.}

It is also important to note that 
filaments in density distribution can be very different from the filaments extracted from velocity channel maps 
(e.g. \citealt{Cl14}). 
In thin channel maps, the latter are likely to be caused by the velocity crowding effect in the velocity space
\citep{LP00}. 
It means that such filaments can still be seen 
even in homogeneous medium
(see \citealt{LY18}).

\section{Summary}

The pervasive magnetized turbulence is responsible for the formation of ubiquitous density filaments in the ISM. 
Corresponding to different turbulence properties in the multi-phase ISM, the resulting filaments have distinctive features. 
Our main results are summarized as follows.

1. The perpendicular mixing motions in Alfv\'{e}nic turbulence give rise to 
filaments aligned with the local magnetic field. 
As Alfv\'{e}nic turbulence dominates the dynamics of compressible MHD turbulence, parallel filaments generally exist in both diffuse media 
and MCs. 
 
2. The axial ratio of parallel filament depends on the anisotropic scaling of Alfv\'{e}nic turbulence and varies in different turbulence regimes. 
Compared with super-Alfv\'{e}nic  turbulence (e.g., the CNM and MCs), 
we expect more elongated filaments in sub-Alfv\'{e}nic  turbulence (e.g., the WNM).

3. The minimum width of parallel filaments is set by the perpendicular neutral-ion decoupling scale in partially ionized media, i.e., 
the transition scale from anisotropic and isotropic turbulence in neutrals. 
In the case of super-Alfv\'{e}nic turbulence, the maximum width of parallel filaments is given by $L_\text{st}$, i.e., the transition scale from 
isotropic to anisotropic turbulence in strongly coupled neutrals and ions.

4. The shock compression in highly supersonic turbulence (e.g., MCs) accounts for the formation of dense filaments perpendicular to magnetic fields. 
The filament width depends on the length scale of the supersonic turbulent flow and the sonic Mach number.

5. When confronting the theoretically predicted filament features 
with observations, one should take into account the observational effects, e.g., limited resolution, projection effect, 
and be cautious when identifying density filaments from observational data.
\\
\\

S.X. acknowledges the support for Program number HST-HF2-51400.001-A provided by NASA through a grant from the Space Telescope Science Institute, which is operated by the Association of Universities for Research in Astronomy, Incorporated, under NASA contract NAS5-26555.
A.L. acknowledges the support from grant NSF DMS 1622353.

\bibliographystyle{apj.bst}
\bibliography{xu}

\begin{thebibliography}{60}
\expandafter\ifx\csname natexlab\endcsname\relax\def\natexlab#1{#1}\fi

\bibitem[{{Andr{\'e}} {et~al.}(2014){Andr{\'e}}, {Di Francesco},
  {Ward-Thompson}, {Inutsuka}, {Pudritz}, \& {Pineda}}]{And14}
{Andr{\'e}}, P., {Di Francesco}, J., {Ward-Thompson}, D., {Inutsuka}, S.-I.,
  {Pudritz}, R.~E., \& {Pineda}, J.~E. 2014, Protostars and Planets VI, 27

\bibitem[{{Andr{\'e}} {et~al.}(2010){Andr{\'e}}, {Men'shchikov}, {Bontemps},
  {K{\"o}nyves}, {Motte}, {Schneider}, {Didelon}, {Minier}, {Saraceno},
  {Ward-Thompson}, {di Francesco}, {White}, {Molinari}, {Testi}, {Abergel},
  {Griffin}, {Henning}, {Royer}, {Mer{\'{\i}}n}, {Vavrek}, {Attard},
  {Arzoumanian}, {Wilson}, {Ade}, {Aussel}, {Baluteau}, {Benedettini},
  {Bernard}, {Blommaert}, {Cambr{\'e}sy}, {Cox}, {di Giorgio}, {Hargrave},
  {Hennemann}, {Huang}, {Kirk}, {Krause}, {Launhardt}, {Leeks}, {Le Pennec},
  {Li}, {Martin}, {Maury}, {Olofsson}, {Omont}, {Peretto}, {Pezzuto}, {Prusti},
  {Roussel}, {Russeil}, {Sauvage}, {Sibthorpe}, {Sicilia-Aguilar}, {Spinoglio},
  {Waelkens}, {Woodcraft}, \& {Zavagno}}]{And10}
{Andr{\'e}}, P., {et~al.} 2010, \aap, 518, L102

\bibitem[{{Armstrong} {et~al.}(1995){Armstrong}, {Rickett}, \&
  {Spangler}}]{Armstrong95}
{Armstrong}, J.~W., {Rickett}, B.~J., \& {Spangler}, S.~R. 1995, \apj, 443, 209

\bibitem[{{Banerjee} {et~al.}(2006){Banerjee}, {Pudritz}, \&
  {Anderson}}]{Ban06}
{Banerjee}, R., {Pudritz}, R.~E., \& {Anderson}, D.~W. 2006, \mnras, 373, 1091

\bibitem[{{Beresnyak} {et~al.}(2005){Beresnyak}, {Lazarian}, \& {Cho}}]{BLC05}
{Beresnyak}, A., {Lazarian}, A., \& {Cho}, J. 2005, \apjl, 624, L93

\bibitem[{{Burkhart} {et~al.}(2015{\natexlab{a}}){Burkhart}, {Collins}, \&
  {Lazarian}}]{Bur15}
{Burkhart}, B., {Collins}, D.~C., \& {Lazarian}, A. 2015{\natexlab{a}}, \apj,
  808, 48

\bibitem[{{Burkhart} {et~al.}(2009){Burkhart}, {Falceta-Gon{\c c}alves},
  {Kowal}, \& {Lazarian}}]{Burk09}
{Burkhart}, B., {Falceta-Gon{\c c}alves}, D., {Kowal}, G., \& {Lazarian}, A.
  2009, \apj, 693, 250

\bibitem[{{Burkhart} {et~al.}(2015{\natexlab{b}}){Burkhart}, {Lazarian},
  {Balsara}, {Meyer}, \& {Cho}}]{Burk15}
{Burkhart}, B., {Lazarian}, A., {Balsara}, D., {Meyer}, C., \& {Cho}, J.
  2015{\natexlab{b}}, \apj, 805, 118

\bibitem[{{Burkhart} {et~al.}(2010){Burkhart}, {Stanimirovi{\'c}}, {Lazarian},
  \& {Kowal}}]{Burk10}
{Burkhart}, B., {Stanimirovi{\'c}}, S., {Lazarian}, A., \& {Kowal}, G. 2010,
  \apj, 708, 1204

\bibitem[{{Chepurnov} \& {Lazarian}(2010)}]{CheL10}
{Chepurnov}, A., \& {Lazarian}, A. 2010, \apj, 710, 853

\bibitem[{{Chepurnov} {et~al.}(2010){Chepurnov}, {Lazarian},
  {Stanimirovi{\'c}}, {Heiles}, \& {Peek}}]{Chep10}
{Chepurnov}, A., {Lazarian}, A., {Stanimirovi{\'c}}, S., {Heiles}, C., \&
  {Peek}, J.~E.~G. 2010, \apj, 714, 1398

\bibitem[{{Cho} \& {Lazarian}(2002)}]{CL02_PRL}
{Cho}, J., \& {Lazarian}, A. 2002, Physical Review Letters, 88, 245001

\bibitem[{{Cho} \& {Lazarian}(2003)}]{CL03}
---. 2003, \mnras, 345, 325

\bibitem[{{Cho} {et~al.}(2002){Cho}, {Lazarian}, \& {Vishniac}}]{CLV_incomp}
{Cho}, J., {Lazarian}, A., \& {Vishniac}, E.~T. 2002, \apj, 564, 291

\bibitem[{{Clark} {et~al.}(2015){Clark}, {Hill}, {Peek}, {Putman}, \&
  {Babler}}]{Cl15}
{Clark}, S.~E., {Hill}, J.~C., {Peek}, J.~E.~G., {Putman}, M.~E., \& {Babler},
  B.~L. 2015, Physical Review Letters, 115, 241302

\bibitem[{{Clark} {et~al.}(2014){Clark}, {Peek}, \& {Putman}}]{Cl14}
{Clark}, S.~E., {Peek}, J.~E.~G., \& {Putman}, M.~E. 2014, \apj, 789, 82

\bibitem[{{Collins} {et~al.}(2012){Collins}, {Kritsuk}, {Padoan}, {Li}, {Xu},
  {Ustyugov}, \& {Norman}}]{Col12}
{Collins}, D.~C., {Kritsuk}, A.~G., {Padoan}, P., {Li}, H., {Xu}, H.,
  {Ustyugov}, S.~D., \& {Norman}, M.~L. 2012, \apj, 750, 13

\bibitem[{{Draine}(2011)}]{Draine:2011aa}
{Draine}, B.~T. 2011, {Physics of the Interstellar and Intergalactic Medium}
  (Princeton University Press)

\bibitem[{{Draine} \& {Lazarian}(1998)}]{Dr98}
{Draine}, B.~T., \& {Lazarian}, A. 1998, \apjl, 494, L19

\bibitem[{{Draine} {et~al.}(1983){Draine}, {Roberge}, \& {Dalgarno}}]{Drai83}
{Draine}, B.~T., {Roberge}, W.~G., \& {Dalgarno}, A. 1983, \apj, 264, 485

\bibitem[{{Federrath} \& {Klessen}(2012)}]{Fed12}
{Federrath}, C., \& {Klessen}, R.~S. 2012, \apj, 761, 156

\bibitem[{{Goldreich} \& {Sridhar}(1995)}]{GS95}
{Goldreich}, P., \& {Sridhar}, S. 1995, \apj, 438, 763

\bibitem[{{Gonz{\'a}lez-Casanova} \& {Lazarian}(2017)}]{Gon17}
{Gonz{\'a}lez-Casanova}, D.~F., \& {Lazarian}, A. 2017, \apj, 835, 41

\bibitem[{{Hennebelle} \& {Falgarone}(2012)}]{HF12}
{Hennebelle}, P., \& {Falgarone}, E. 2012, \aapr, 20, 55

\bibitem[{{Kowal} {et~al.}(2007){Kowal}, {Lazarian}, \& {Beresnyak}}]{KL07}
{Kowal}, G., {Lazarian}, A., \& {Beresnyak}, A. 2007, \apj, 658, 423

\bibitem[{{Kowal} {et~al.}(2009){Kowal}, {Lazarian}, {Vishniac}, \&
  {Otmianowska-Mazur}}]{KL09}
{Kowal}, G., {Lazarian}, A., {Vishniac}, E.~T., \& {Otmianowska-Mazur}, K.
  2009, \apj, 700, 63

\bibitem[{{Kowal} {et~al.}(2012){Kowal}, {Lazarian}, {Vishniac}, \&
  {Otmianowska-Mazur}}]{KL12}
---. 2012, Nonlinear Processes in Geophysics, 19, 297

\bibitem[{{Lada} {et~al.}(2008){Lada}, {Muench}, {Rathborne}, {Alves}, \&
  {Lombardi}}]{Lad08}
{Lada}, C.~J., {Muench}, A.~A., {Rathborne}, J., {Alves}, J.~F., \& {Lombardi},
  M. 2008, \apj, 672, 410

\bibitem[{{Larson}(1981)}]{Lars81}
{Larson}, R.~B. 1981, \mnras, 194, 809

\bibitem[{{Lazarian}(2005)}]{Laz05}
{Lazarian}, A. 2005, in American Institute of Physics Conference Series, Vol.
  784, Magnetic Fields in the Universe: From Laboratory and Stars to Primordial
  Structures., ed. E.~M. {de Gouveia dal Pino}, G.~{Lugones}, \& A.~{Lazarian},
  42--53

\bibitem[{{Lazarian}(2006)}]{Lazarian06}
{Lazarian}, A. 2006, \apjl, 645, L25

\bibitem[{{Lazarian}(2009)}]{Laz09rev}
---. 2009, Space Science Reviews, 143, 357

\bibitem[{{Lazarian} {et~al.}(2012){Lazarian}, {Esquivel}, \&
  {Crutcher}}]{LEC12}
{Lazarian}, A., {Esquivel}, A., \& {Crutcher}, R. 2012, \apj, 757, 154

\bibitem[{{Lazarian} \& {Pogosyan}(2000)}]{LP00}
{Lazarian}, A., \& {Pogosyan}, D. 2000, \apj, 537, 720

\bibitem[{{Lazarian} \& {Vishniac}(1999)}]{LV99}
{Lazarian}, A., \& {Vishniac}, E.~T. 1999, \apj, 517, 700

\bibitem[{{Lazarian} {et~al.}(2004){Lazarian}, {Vishniac}, \& {Cho}}]{LVC04}
{Lazarian}, A., {Vishniac}, E.~T., \& {Cho}, J. 2004, \apj, 603, 180

\bibitem[{{Lazarian} \& {Yuen}(2017)}]{LY18}
{Lazarian}, A., \& {Yuen}, K.~H. 2017, ArXiv: 1703.03119

\bibitem[{{Lithwick} \& {Goldreich}(2001)}]{LG01}
{Lithwick}, Y., \& {Goldreich}, P. 2001, \apj, 562, 279

\bibitem[{{Maron} \& {Goldreich}(2001)}]{MG01}
{Maron}, J., \& {Goldreich}, P. 2001, \apj, 554, 1175

\bibitem[{{Marsh} {et~al.}(2016){Marsh}, {Kirk}, {Andr{\'e}}, {Griffin},
  {K{\"o}nyves}, {Palmeirim}, {Men'shchikov}, {Ward-Thompson}, {Benedettini},
  {Bresnahan}, {di Francesco}, {Elia}, {Motte}, {Peretto}, {Pezzuto}, {Roy},
  {Sadavoy}, {Schneider}, {Spinoglio}, \& {White}}]{Mar16}
{Marsh}, K.~A., {et~al.} 2016, \mnras, 459, 342

\bibitem[{{McClure-Griffiths} {et~al.}(2006){McClure-Griffiths}, {Dickey},
  {Gaensler}, {Green}, \& {Haverkorn}}]{McC06}
{McClure-Griffiths}, N.~M., {Dickey}, J.~M., {Gaensler}, B.~M., {Green}, A.~J.,
  \& {Haverkorn}, M. 2006, \apj, 652, 1339

\bibitem[{{McKee} \& {Ostriker}(2007)}]{Mckee_Ostriker2007}
{McKee}, C.~F., \& {Ostriker}, E.~C. 2007, \araa, 45, 565

\bibitem[{{Pan} \& {Scannapieco}(2010)}]{Pan10}
{Pan}, L., \& {Scannapieco}, E. 2010, \apj, 721, 1765

\bibitem[{{Planck Collaboration} {et~al.}(2016{\natexlab{a}}){Planck
  Collaboration}, {Adam}, {Ade}, {Aghanim}, {Alves}, {Arnaud}, {Arzoumanian},
  {Ashdown}, {Aumont}, {Baccigalupi}, \& et~al.}]{Pla16}
{Planck Collaboration} {et~al.} 2016{\natexlab{a}}, \aap, 586, A135

\bibitem[{{Planck Collaboration} {et~al.}(2016{\natexlab{b}}){Planck
  Collaboration}, {Ade}, {Aghanim}, {Alves}, {Arnaud}, {Arzoumanian},
  {Ashdown}, {Aumont}, {Baccigalupi}, {Banday}, {Barreiro}, {Bartolo},
  {Battaner}, {Benabed}, {Beno{\^\i}t}, {Benoit-L{\'e}vy}, {Bernard},
  {Bersanelli}, {Bielewicz}, {Bock}, {Bonavera}, {Bond}, {Borrill}, {Bouchet},
  {Boulanger}, {Bracco}, {Burigana}, {Calabrese}, {Cardoso}, {Catalano},
  {Chiang}, {Christensen}, {Colombo}, {Combet}, {Couchot}, {Crill}, {Curto},
  {Cuttaia}, {Danese}, {Davies}, {Davis}, {de Bernardis}, {de Rosa}, {de
  Zotti}, {Delabrouille}, {Dickinson}, {Diego}, {Dole}, {Donzelli}, {Dor{\'e}},
  {Douspis}, {Ducout}, {Dupac}, {Efstathiou}, {Elsner}, {En{\ss}lin},
  {Eriksen}, {Falceta-Gon{\c c}alves}, {Falgarone}, {Ferri{\`e}re}, {Finelli},
  {Forni}, {Frailis}, {Fraisse}, {Franceschi}, {Frejsel}, {Galeotta}, {Galli},
  {Ganga}, {Ghosh}, {Giard}, {Gjerl{\o}w}, {Gonz{\'a}lez-Nuevo}, {G{\'o}rski},
  {Gregorio}, {Gruppuso}, {Gudmundsson}, {Guillet}, {Harrison}, {Helou},
  {Hennebelle}, {Henrot-Versill{\'e}}, {Hern{\'a}ndez-Monteagudo}, {Herranz},
  {Hildebrandt}, {Hivon}, {Holmes}, {Hornstrup}, {Huffenberger}, {Hurier},
  {Jaffe}, {Jaffe}, {Jones}, {Juvela}, {Keih{\"a}nen}, {Keskitalo}, {Kisner},
  {Knoche}, {Kunz}, {Kurki-Suonio}, {Lagache}, {Lamarre}, {Lasenby},
  {Lattanzi}, {Lawrence}, {Leonardi}, {Levrier}, {Liguori}, {Lilje},
  {Linden-V{\o}rnle}, {L{\'o}pez-Caniego}, {Lubin}, {Mac{\'{\i}}as-P{\'e}rez},
  {Maino}, {Mandolesi}, {Mangilli}, {Maris}, {Martin},
  {Mart{\'{\i}}nez-Gonz{\'a}lez}, {Masi}, {Matarrese}, {Melchiorri}, {Mendes},
  {Mennella}, {Migliaccio}, {Miville-Desch{\^e}nes}, {Moneti}, {Montier},
  {Morgante}, {Mortlock}, {Munshi}, {Murphy}, {Naselsky}, {Nati},
  {Netterfield}, {Noviello}, {Novikov}, {Novikov}, {Oppermann}, {Oxborrow},
  {Pagano}, {Pajot}, {Paladini}, {Paoletti}, {Pasian}, {Perotto}, {Pettorino},
  {Piacentini}, {Piat}, {Pierpaoli}, {Pietrobon}, {Plaszczynski},
  {Pointecouteau}, {Polenta}, {Ponthieu}, {Pratt}, {Prunet}, {Puget}, {Rachen},
  {Reinecke}, {Remazeilles}, {Renault}, {Renzi}, {Ristorcelli}, {Rocha},
  {Rossetti}, {Roudier}, {Rubi{\~n}o-Mart{\'{\i}}n}, {Rusholme}, {Sandri},
  {Santos}, {Savelainen}, {Savini}, {Scott}, {Soler}, {Stolyarov}, {Sudiwala},
  {Sutton}, {Suur-Uski}, {Sygnet}, {Tauber}, {Terenzi}, {Toffolatti}, {Tomasi},
  {Tristram}, {Tucci}, {Umana}, {Valenziano}, {Valiviita}, {Van Tent},
  {Vielva}, {Villa}, {Wade}, {Wandelt}, {Wehus}, {Ysard}, {Yvon}, \&
  {Zonca}}]{Pln16}
---. 2016{\natexlab{b}}, \aap, 586, A138

\bibitem[{{Pudritz} \& {Kevlahan}(2013)}]{Pud13}
{Pudritz}, R.~E., \& {Kevlahan}, N.~K.-R. 2013, Philosophical Transactions of
  the Royal Society of London Series A, 371, 20120248

\bibitem[{{Santos-Lima} {et~al.}(2010){Santos-Lima}, {Lazarian}, {de Gouveia
  Dal Pino}, \& {Cho}}]{Sant10}
{Santos-Lima}, R., {Lazarian}, A., {de Gouveia Dal Pino}, E.~M., \& {Cho}, J.
  2010, \apj, 714, 442

\bibitem[{{Schneider} \& {Elmegreen}(1979)}]{SE79}
{Schneider}, S., \& {Elmegreen}, B.~G. 1979, \apjs, 41, 87

\bibitem[{{Shu}(1992)}]{Shu92}
{Shu}, F.~H. 1992, {The physics of astrophysics. Volume II: Gas dynamics.}

\bibitem[{{Whitworth} \& {Clarke}(1997)}]{Wh97}
{Whitworth}, A.~P., \& {Clarke}, C.~J. 1997, \mnras, 291, 578

\bibitem[{{Williams} {et~al.}(2000){Williams}, {Blitz}, \& {McKee}}]{Will00}
{Williams}, J.~P., {Blitz}, L., \& {McKee}, C.~F. 2000, Protostars and Planets
  IV, 97

\bibitem[{{Xu} {et~al.}(2015){Xu}, {Lazarian}, \& {Yan}}]{XuL15}
{Xu}, S., {Lazarian}, A., \& {Yan}, H. 2015, \apj, 810, 44

\bibitem[{{Xu} {et~al.}(2016){Xu}, {Yan}, \& {Lazarian}}]{Xuc16}
{Xu}, S., {Yan}, H., \& {Lazarian}, A. 2016, \apj, 826, 166

\bibitem[{{Xu} \& {Zhang}(2017{\natexlab{a}})}]{XZg17}
{Xu}, S., \& {Zhang}, B. 2017{\natexlab{a}}, \apjl, 846, L28

\bibitem[{{Xu} \& {Zhang}(2017{\natexlab{b}})}]{XuZ17}
---. 2017{\natexlab{b}}, \apj, 835, 2

\bibitem[{{Yan} \& {Lazarian}(2002)}]{YL02}
{Yan}, H., \& {Lazarian}, A. 2002, Physical Review Letters, 89, B1102+

\bibitem[{{Yuen} \& {Lazarian}(2017{\natexlab{a}})}]{YuL17}
{Yuen}, K.~H., \& {Lazarian}, A. 2017{\natexlab{a}}, arXiv:1703.03026

\bibitem[{{Yuen} \& {Lazarian}(2017{\natexlab{b}})}]{Yu17}
---. 2017{\natexlab{b}}, \apjl, 837, L24

\bibitem[{{Zuckerman} \& {Evans}(1974)}]{Zu74}
{Zuckerman}, B., \& {Evans}, II, N.~J. 1974, \apjl, 192, L149

\bibitem[{{Zuckerman} \& {Palmer}(1974)}]{Zuck74}
{Zuckerman}, B., \& {Palmer}, P. 1974, \araa, 12, 279

\end{thebibliography}

\end{document}